\documentclass{aa}
\usepackage{graphicx}
\usepackage{txfonts}
\usepackage{hyperref}
\usepackage{xcolor}
\usepackage{ulem}

\begin{document}

\title{
   Impact of tides on non-coplanar orbits of progenitors of high-mass X-ray binaries
}

%\subtitle{I.}

\author{
    A. Simaz Bunzel\thanks{Fellow of CONICET}
    \inst{1,2}
    \and
    F. Garc\'ia
    \inst{1,2}
    \and
    J. A. Combi
    \inst{1,2,3}
    \and
    F. Fortin
    \inst{4}
    \and
    S. Chaty
    \inst{4}
}

\institute{
    Instituto Argentino de Radioastronom\'ia (CCT La Plata, CONICET; CICPBA; UNLP), C.C.5, (1894) Villa Elisa, Buenos Aires, Argentina
    \and
    Facultad de Ciencias Astron\'omicas y Geof\'{\i}sicas, Universidad Nacional de La Plata, Paseo del Bosque, B1900FWA La Plata, Argentina
    \and
    Departamento de Ingeniería Mecánica y Minera (EPSJ), Universidad de Jaén, Campus Las Lagunillas s/n Ed. A3, E-23071 Jaén, Spain
    \and
    Universit\'e Paris Cit\'e, CNRS, Astroparticule et Cosmologie, F-75013 Paris, France
}

\date{
    received ... 2022 ; accepted ... 2022
}

% \abstract{}{}{}{}{} 
% 5 {} token are mandatory
 
\abstract
    % context heading (optional); {} leave it empty if necessary
    {An important stage in the evolution of massive binaries is the formation of a compact object in the system. It is believed that in
    some cases a momentum kick is  imparted to the newly born object, changing the orbital parameters of the binary, such as
    eccentricity and orbital period, and even acquiring an asynchronous orbit between its components. In this situation, tides play a
    central role in the evolution of these binaries.}
    {In this work we aim to study how the orbital parameters of a massive binary change after the formation of a compact object when the
    stellar spin of the non-degenerate companion is not aligned with the orbital angular momentum.}
    {We used {\tt MESA}, which we modified to be able to evolve binaries with different values of the inclination between the orbital
    planes before and just after the formation of the compact object. These modifications to the equations solved by the {\tt MESA} code
    are extended to the case of non-solid body rotation.}
    {We find that the impact of having different initial inclinations is mostly present in the evolution towards an equilibrium state that
    is independent of the inclination. If the binary separation is small enough such that the interaction happens when the star is burning
    hydrogen in its core, this state is reached before the beginning of a mass-transfer phase, while for a wider binary not all conditions
    characterizing the equilibrium are met. We also explore the effect of having different initial rotation rates in the stars and how the
    Spruit-Tayler dynamo mechanism affects the angular momentum transport for a non-coplanar binary.}
    {These findings show that including the inclination in the equations of tidal evolution to a binary after a kick is imparted onto a
    newly born compact object changes the evolution of some parameters, such as the eccentricity and the spin period of the star, depending
    on how large this inclination is. Moreover, these results can be used to match the properties of observed X-ray binaries to estimate
    the strength of the momentum kick.}

\keywords{
    (stars:) binaries (including multiple): close -- stars: evolution -- X-rays: binaries
}

\maketitle

%-------------------------------------------------------------------
\section{Introduction}
\label{section:introduction}

Tidal evolution operates in a binary system by changing both orbital and rotational parameters of its components. In a closed binary
system, the conservation of angular momentum (AM) leads to an exchange of AM between the orbit and the stars, which tends to an equilibrium
state characterized by a circular orbit (circularization), with stars rotating at a rate equal to the orbital motion (synchronous orbit),
and with their spin axes parallel to the orbital AM (coplanarity). Moreover, tides are tightly correlated to the separation in the binary
such that the closer the components of a binary are, the stronger the tides acting on it will be, thus causing faster evolution to the equilibrium
state. Furthermore, the strength of the tidal interaction depends on the physical process responsible for the dissipation of kinetic energy
\citep{zahn2008}. In reality, binaries are not closed systems and a sink of AM is present in the form of stellar winds, gravitational
radiation, and even inefficient mass-transfer \citep{landau1975,soberman1997}, so that binaries can only attain equilibrium conditions for a
brief period of time.

In particular, the evolution of tides is of great importance for the progenitors of X-ray binaries (XRBs), composed by a non-degenerate
star interacting with a compact object, and for double compact object (DCO) systems. The compact object, which can either be  neutron
star (NS) or a black hole (BH), is formed at the end of the evolution of a massive star due to the unavoidable collapse of its outer
layers onto its central regions, driven by the overwhelming gravitational attraction \citep{hoyle1960,janka2012}. During this stage some
ejection of mass can naturally impart a momentum kick to the newly formed compact object obtained from linear momentum conservation
\citep{blaauw1961}. Furthermore, an asymmetry present during the collapse phase can also lead to a change in the orbital parameters
\citep{kalogera1996}. The origin of these asymmetries may be related to either neutrino emission
\citep[][and references therein]{scheck2006} or to mass ejection during the supernova (SN) caused by rotation in the collapsing core
\citep{kotake2003}, hydrodynamic instabilities, magnetic fields, or convective motions \citep{herant1994, burrows1995, janka1996, keil1996}.
The necessity of this asymmetry lies in observations of young isolated pulsars with space velocities as high as $400$~km~s$^{-1}$
\citep{hobbs2005},     too high to have been born from the breakup of a binary system in a SN explosion.
After these changes take place in a binary, the new   configuration will be out of equilibrium and the orbit will become eccentric
in addition to having some degree of asynchronous orbit between the components. It is in these cases that tides will operate to find a new
equilibrium condition.
% Moreover, \citet{kalogera1998} show that without asymmetric kicks the formation of low-mass XRBs (LMXBs) with short orbital periods
% cannot be explained.

On the one hand, tides are believed to replenish the AM of a star, spinning it up and potentially becoming able to produce rapidly rotating
compact objects \citep{qin2019,belczynski2020,fishbach2021}. There is a hint that this mechanism operates on the observed population of
high-mass XRBs (HMXBs) containing a BH, as measurements of BH spins show high values \citep{miller2015}, although these measurements relay
on model-dependent fits \citep{belczynski2021}. It is worth mentioning that  \citet{qin2022a} have recently found that allowing for a
regime of hypercritical accretion can  explain at least one case of a rapidly rotating BH in a HMXB. On the other hand, most of the
spins of BHs measured by the LIGO and Virgo collaboration \citep{abbott2021} are small. It is believed that in these cases the tides are not
strong enough to cause a spin up of the stars and an efficient AM transport leads to slowly rotating compact objects
\citep{qin2018,olejak2021,qin2022b}. Moreover, some models for long gamma-ray bursts require rapidly rotating massive stars at the end of
their evolution \citep{woosley1993,paczynski1998}, which could be achieved thanks to the spin up of the progenitor stars due to the action
of tides. Thus, it is clear that the interplay between tides and AM transport takes a fundamental role in shaping the different outcomes of
the evolution of massive binaries containing compact objects.

Most of the works in which tidal evolution is implicitly considered assume that coplanarity is instantly achieved, and only follow
numerically the evolution towards a circular and synchronous orbit. This approach is based on the fact that the timescales for each
equilibrium condition are different to each other, with   coplanarity being reached faster than the rest. However, a difference
might still be present when considering the effect of non-zero inclinations in the equations of tidal evolution. In this work, our  aim is to
understand the role that non-coplanarity plays in the evolution of different binary parameters of the immediate progenitors of HMXBs. In
order to do so, we consider the set of differential equations of tidal evolution   derived by \citet{repetto2014}, which we incorporate in
the publicly available stellar evolution code {\tt MESA}. Before using it to  analyse the evolution of HMXBs, we compare our code with the
outcome of other codes in the literature. The paper is organized  as follows: in Section~\ref{section:model} we describe the model used
together with the additions included in the stellar evolution code. In Section~\ref{section:results} we present the results of some
exemplary cases of study. We present a summary of our findings and concluding remarks in Section~\ref{section:conclusions}.

%--------------------------------------------------------------------
\section{Methods}
\label{section:model}

\subsection{Tidal modelling}
\label{subsection:tidal-model}

Orbital parameters can be modified as a consequence of tides operating in a binary. These tidal forces, which are present in the surface of
a stellar component, will not be aligned with a line joining the centre of the components of the binary, due to different mechanisms of
energy dissipation, thus producing a torque in the system. Then, thanks to the spin-orbit coupling, tidal interactions can modify the
orbital AM as well as the spin of the components of the binary. Hence, to understand how these tides operate in a binary, it is important
to have a model to account for their evolution.

In order to model the effect of tides acting on a non-coplanar progenitor of a HMXB, we consider the equations presented in
\citet{repetto2014} which are valid for arbitrary inclinations between the stellar spin and the orbital AM. That work was an extension of
the tidal-interaction study presented in \citet{hut1981}, where only small-angle deviations were considered. In this work we combine those
equations with the adjustments described in \citet{paxton2015} to include the case of differentially rotating stars, in which the AM is
distributed throughout the entire stellar interior and envelope.

This particular model of tidal evolution, where only tides with a small deviation from the equilibrium shape are considered, is
called the weak friction model \citep[see e.g.][]{darwin1879,alexander1973}. Here it is assumed that the deviation of tides in
magnitude and direction are parametrized by a constant time lag, thus allowing for slow variations of orbital parameters within an orbital
period. Dynamical effects, in which stars can oscillate, are not considered. For a review of these dynamical effects, see \citet{zahn1977}.

Coupled with the stellar evolution of the binary, the tidal equations to solve are

\begin{equation}
   \begin{aligned}
      \dfrac{{\rm d}a}{{\rm d}t} = &-6 \, \left( \dfrac{K}{T} \right)_1 \, q (1 + q) \, \left( \dfrac{R_1}{a} \right)^8 \, \dfrac{a}{(1 - e^2)^{15/2}} \, \Bigg[ f_1(e^2) \\
                                   & - (1 - e^2)^{3/2} \, f_2(e^2) \, \dfrac{\Omega_1 \, \cos i}{\Omega_{\rm orb}} \Bigg],
   \end{aligned}
\end{equation}

\begin{equation}
   \begin{aligned}
      \dfrac{{\rm d}e}{{\rm d}t} = &-27 \, \left( \dfrac{K}{T} \right)_1 \, q (1 + q) \, \left( \dfrac{R_1}{a} \right)^8 \, \dfrac{e}{(1 - e^2)^{13/2}} \, \Bigg[ f_3(e^2) \\
                                   &- \dfrac{11}{8} \, (1 - e^2)^{3/2} \, f_4(e^2) \, \dfrac{\Omega_1 \, \cos i}{\Omega_{\rm orb}} \Bigg],
   \end{aligned}
\end{equation}

\begin{equation}
   \begin{aligned}
      \dfrac{{\rm d}\Omega_1}{{\rm d}t} = &-3 \, \left( \dfrac{K}{T} \right)_1 \, \dfrac{q^2}{k^2} \, \left( \dfrac{R_1}{a} \right)^6 \, \dfrac{\Omega_{\rm orb}}{(1 - e^2)^{6}}
      \, \Bigg[
      f_2(e^2) \, \cos i - \dfrac{\Omega_1}{4 \, \Omega_{\rm orb}} \\
                                        &(1 - e^2)^{3/2} \, (3 + 2 \cos 2i) \, f_5(e^2) \, \dfrac{\Omega_1 \, \cos i}{\Omega_{\rm orb}} \Bigg],
   \end{aligned}
\end{equation}

\begin{equation}\label{inclination-equation}
   \begin{aligned}
      \dfrac{{\rm d}i}{{\rm d}t} = &-3 \, \left( \dfrac{K}{T} \right)_1 \, \dfrac{q^2}{k^2} \, \left( \dfrac{R_1}{a} \right)^6 \, \dfrac{\Omega_{\rm orb} \sin i}{\Omega_1 \, (1 -
      e^2)^{6}} \, \Bigg[
      f_2(e^2) - \dfrac{f_5(e^2)}{2} \\
                                   &\times \, \left( \dfrac{\Omega_1 \, \cos i \, (1-e^2)^{3/2}}{\Omega_{\rm orb}} + \dfrac{R_1^2 \, a \, \Omega_1^2 \, k_1^2 \, (1-e^2)}{M_2 \, G}
                                \right) \Bigg].
   \end{aligned}
\end{equation}Here the sub-indices~1~and~2 refer to the non-degenerate star and the compact companion, respectively. The term $(K/T)_1$ is a measure of
the strength of the tidal dissipation, with $K$ being the apsidal motion constant and $T$ the timescale at which changes in the orbit occur
due to tides. The mass ratio~($q$) is defined as $q = M_2/M_1$; $k$ is the radius of gyration of the star, equal to $k = I / (M\,R)^2$ with
$I$ being the moment of inertia of the star; and $f_i(e^2)$ are the polynomials of the eccentricity, $e$, as presented in \citet{hut1981}; and
$\Omega_1$ is the rotational velocity of the star, while $\Omega_{\rm orb}$ is the orbital angular velocity. The equations show the strong
dependence with the orbital separation ($a$), coupled with the radius of the star ($R_1$), as well as the role of the inclination ($i$)
between the orbital and the star angular momenta.

The term $K/T$ depends on the source of dissipation of kinetic energy, in which $T$ is a characteristic timescale for that source of
dissipation. For stars with a convective envelope, this source is the turbulent convection, so that $T$ is the eddy turnover timescale. For
this type of stars we consider a calibration factor of

\begin{equation}
   \left( \dfrac{K}{T} \right)_1 = \dfrac{2}{21} \dfrac{f_{\rm conv}}{\tau_{\rm conv}} \dfrac{M_{\rm env}}{M_1}
,\end{equation}

{\noindent where $f_{\rm conv}$ is the fraction of convective cells that contribute to the damping, for which we assume a quadratic
dependence with the tidal-forcing frequency \citep{goldreich1977}, increased by a factor of 50 \citep{belczynski2008}.}

In the case of a star with a radiative envelope, dissipation is assumed to be produced by radiative damping. We use the calibration factor
as in \citet{hurley2002}

\begin{equation}
   \left( \dfrac{K}{T} \right)_1 = 1.9782 \times 10^4 \, \dfrac{M_1 \, R_1^2}{a^5} \, (1+q)^{5/6} \, E_2
,\end{equation}

{\noindent where $E_2$ measures the coupling between the tidal potential and gravity mode as given by \citet{zahn1975}.}

Although the agreement between the observations and the theoretical predictions of tidal torques is quite satisfactory, the complex
physical mechanisms that are at the origin of the tidal torques are not completely understood so that the evaluation of $K/T$ require some
calibration factors, as mentioned before. Naturally, given the physical origins of the convective turbulence and the radiative damping,
their efficiencies are different with convective envelope stars presenting more tidal friction than those with radiative envelopes.

Then we integrate the set of equations for tidal evolution described above and couple them with the evolution of the orbital AM. Once the
value of $i$ is known at a given timestep from solving Equation~\ref{inclination-equation}, we use the hooks provided by the {\tt MESA}
code for the following: the {\tt other\_edot\_tidal} subroutine is used to solve the evolution of $e$, while the
{\tt other\_sync\_spin\_to\_orbit} subroutine computes the tidal torque acting on the non-degenerate companion. \footnote{The modified
version of the code can be found in the following GitHub repository: \url{https://github.com/asimazbunzel/non-coplanar-tides}}

\subsection{Stellar modelling}
\label{subsection:stellar-model}

The detailed evolution of binary systems is computed using the stellar evolution code {\tt MESA}
\citep[version 15140,][]{paxton2011, paxton2013, paxton2015, paxton2018, paxton2019}. All computed stars are assumed to have a solar
metallicity content, $Z = Z_\odot = 0.017$ \citep{grevesse1998}. We use the  default nuclear reactions networks present in {\tt MESA},
{\tt basic.net}, {\tt co\_burn.net} and {\tt approx21.net}, which are switched dynamically as later stages in the evolution are reached.
Convective regions, which are defined according to the Ledoux criterion \citep{ledoux1947}, are modelled using the standard mixing-length
theory \citep{bohmvitense1958} with a mixing-length parameter $\alpha_{\rm MLT} = 1.5$. Semi-convection is modelled following
\citet{langer1983} with an efficient parameter $\alpha_{\rm SC} = 1$. We include convective core overshooting during core hydrogen burning
following \citet{brott2011}. In order to avoid some numerical problems during the evolution of massive stars until the Wolf-Rayet~(WR)
stage, we use the MLT++ formalism, as presented in \citet{paxton2013}, which reduces the super-adiabaticity in regions where the convective
velocities approach the speed of sound. Additionally, the effect of thermohaline mixing, which accounts for the unstable situation arising
when material accreted by the accretor has a mean molecular weight higher than its outer layers, follows \citet{kippenhahn1980} with an
efficiency parameter of $\alpha_{\rm th} = 1$. The modelling of stellar winds follows that of \citet{brott2011}, as described in
\citet{marchant2017}. Stellar rotation is modelled following the description of \citet{heger2000}, with composition and angular momentum
transport mechanisms including the Eddington-Sweet circulation, secular and dynamical shear instabilities, and the
Goldreich-Schubert-Fricke (GSF) instability. Moreover, the impact of magnetic fields on the transport of AM follows the  \citet{petrovic2005}
implementation, where the Spruit-Tayler (ST) dynamo is taken into account \citep{spruit2002}. Whether this dynamo effect should be
considered in the transport of AM is the object of a current debate:  \citet{qin2019} have shown that without it models of BHs in HMXBs can
reach near critical rotation (spin parameter $a \sim 1$), while models considering its impact in the AM transport leads to BHs with near
zero spin values, in contradiction to spin measurements of BHs in HMXBs. In this work we consider models with and without this mechanism
operating in the AM transport.

In our models a star is initially assumed to be rotating as a solid body. We explore two different cases for the initial rotation rate:
one where the star is rotating at $90\%$ of its critical value, and the other where the star is rotating at $10\%$ of this value. The difference between these two
conditions lies in  how AM is exchanged between the orbit and the star during the evolution. Moreover, we consider AM losses in the orbit
from gravitational wave radiation, mass loss, and spin-orbit coupling, as described in \citet{paxton2015}.

At the start of the calculations, binaries are initially assumed to have already gone through the formation of the compact object, such
that some eccentricity is present. In addition, we also consider the binary to be non-coplanar and with asynchronous orbit between the star
and the compact object. We also neglect the evolution prior to this point of the interior of the companion star to the compact object,
considering that it is represented by a zero age main sequence (ZAMS) star. Binary interaction is computed with the {\tt MESAbinary} module
of {\tt MESA}. When the star in the binary overflows its Roche lobe, the mass transfer (MT) rate is implicitly computed following the
prescription of \citet{kolb1990}. A key factor in the Roche model is the assumption of synchronous orbit, which in our case is not imposed.
As we show, in some of our simulations there are cases where the binary is close to a MT stage, but  have not yet achieved a
synchronous state. Moreover, the model of \citet{hut1981} considers only quadrupolar terms as a perturbation to a central force field,
which for a star nearly filling its Roche lobe might not be the best representation. In this case the tidal coupling should be stronger than
 predicted. For simplicity, we extend the conditions of the Roche model to asynchronous orbit binaries. However, we caution that
this is formally not correct.

In all the cases explored, our simulations are followed until the non-degenerate star reaches the end of the core helium burning phase.

\subsection{Validation of the method}

Before presenting the results for the progenitors of HMXBs, we computed similar cases to other results present in the modern literature
concerning the effect of tides on the evolution of non-coplanar binaries with the aim of verifying and validating our implementation. To this end,
we computed a set of evolutionary models as in \citet{rodriguez2021} as a comparison. In order to do so, we modified the method described in
Section~\ref{subsection:tidal-model} to consider the change over time of each of the components of the binary, including both the
rotational angular velocity and the inclination of the spin with respect to the orbital AM.

The chosen binary is composed of a $17.9$~M$_\odot$ secondary star orbiting around a $33.8$~M$_\odot$ primary star, in eccentric
orbits with different initial periods, $P_{\rm orb}$. We computed three models, as \citet{rodriguez2021} did. These evolutionary
tracks were followed until the end of the main sequence (MS) phase.

\begin{figure}
    \centering
    \includegraphics[width=\hsize]{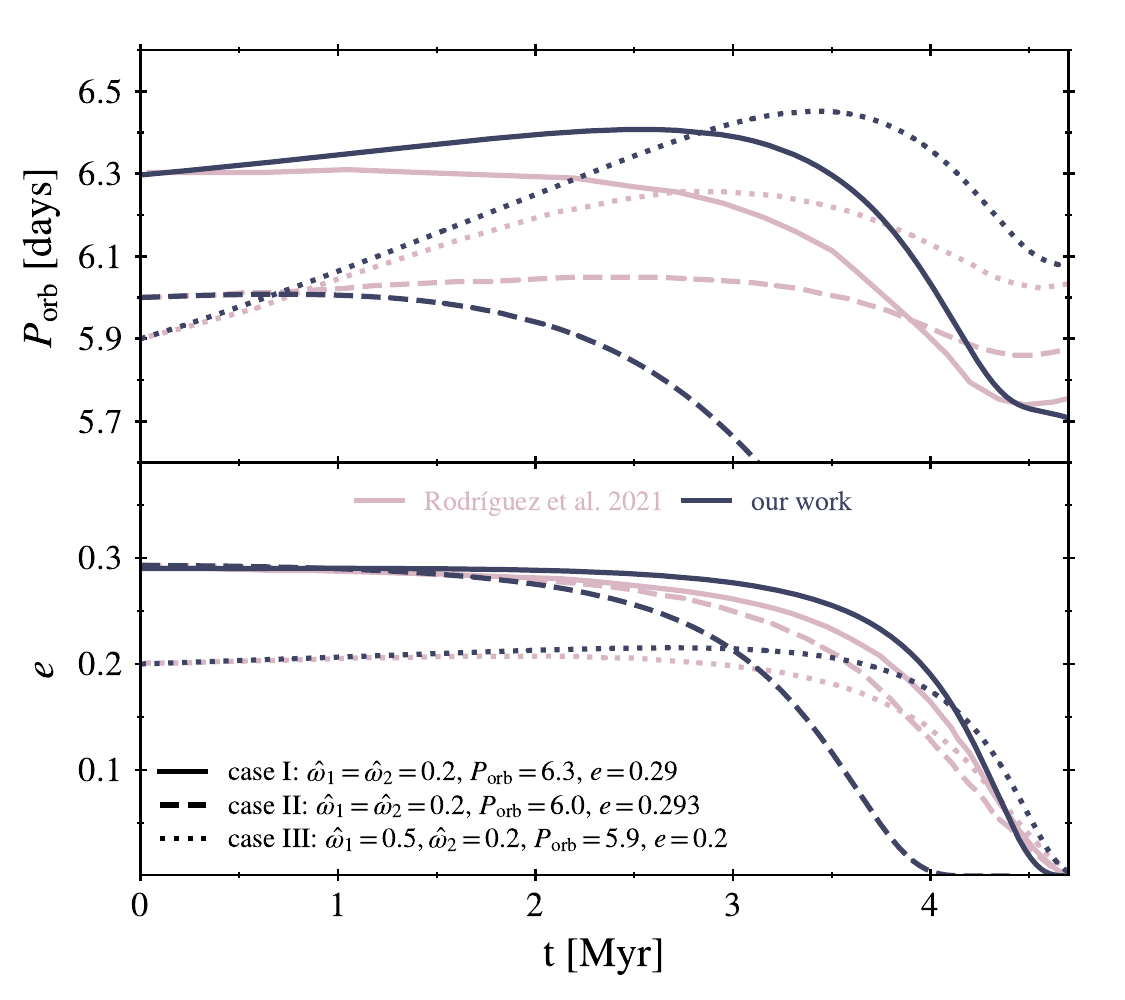}
    \caption{Comparison of the evolution of the orbital parameters for the different scenarios explored by \citet{rodriguez2021}. Shown in dark
       blue are the results found with our modifications to the {\tt MESA} source code, while in  grey  the results from
       \citet{rodriguez2021}. In the top (bottom) panel the evolution of $P_{\rm orb}$ ($e$) is represented for different initial
       conditions (see legend). All three cases  are described in more detail in   
       \citet{rodriguez2021}.}
    \label{fig:rodriguez_comparison}
\end{figure}

In Fig.~\ref{fig:rodriguez_comparison} the evolution of $P_{\rm orb}$ and $e$ are presented. For each individual case covering different
initial conditions, we are able to find similar behaviours in the overall evolution of the orbital parameters. The main differences are
found in the evolution of $P_{\rm orb}$ in the range $1-4$~Myr, which could be associated with the different mechanisms of AM losses
considered in our model. Around 4~Myr the evolutionary tracks for cases I and III seem to behave in the same manner;  during this phase
the stars are close to each other, with tides playing an important role in the evolution.

In contrast, for case II we obtain a faster declining $P_{\rm orb}$ evolution. Compared to that work, this deviation starts at around 2~Myr
and grows over time. Close to 2.5~Myr the binary achieves coplanarity, while synchronous orbit is reached at 3.8~Myr. Finally, at about
4~Myr the orbit   becomes circular. However, our value of $P_{\rm orb}$ is lower than that found by \citet{rodriguez2021}, likely due to differences
in how AM is computed in our code with respect to theirs.

In addition, in the bottom panel of Fig.~\ref{fig:rodriguez_comparison} we show that the evolution of $e$ closely resembles the results
of \citet{rodriguez2021} in all the cases. Once again, in case II we find that our simulation evolves towards a circular orbit  
faster than theirs.

%--------------------------------------------------------------------
\section{Results}
\label{section:results}

In this section, we show the results on the evolution of different binary parameters subject to non-coplanar tides.

\subsection{Mass transfer during core hydrogen burning}
\label{subsection:results-caseA}

First, we model the evolution of a progenitor of a HMXB in which the MT between the star and the compact object occurs when the
non-degenerate star is in the MS, called case A of MT \citep{kippenhahn1967}. For this, we model a binary consisting of a
$15$~M$_\odot$ BH in an eccentric orbit with $e = 0.25$ and $P_{\rm orb} = 3$~days, around a $20$~M$_\odot$ companion star. The star is
assumed to be rotating at $0.9$ of its critical rotation rate ($\Omega_1 / \Omega_{\rm crit} = 0.9$). To understand the role of
non-coplanarity, we explore several different  inclinations between the spin of the star and the orbital AM, as well as the coplanar case
of ($i = 0$).

\begin{figure}
   \centering
   \includegraphics[width=\hsize]{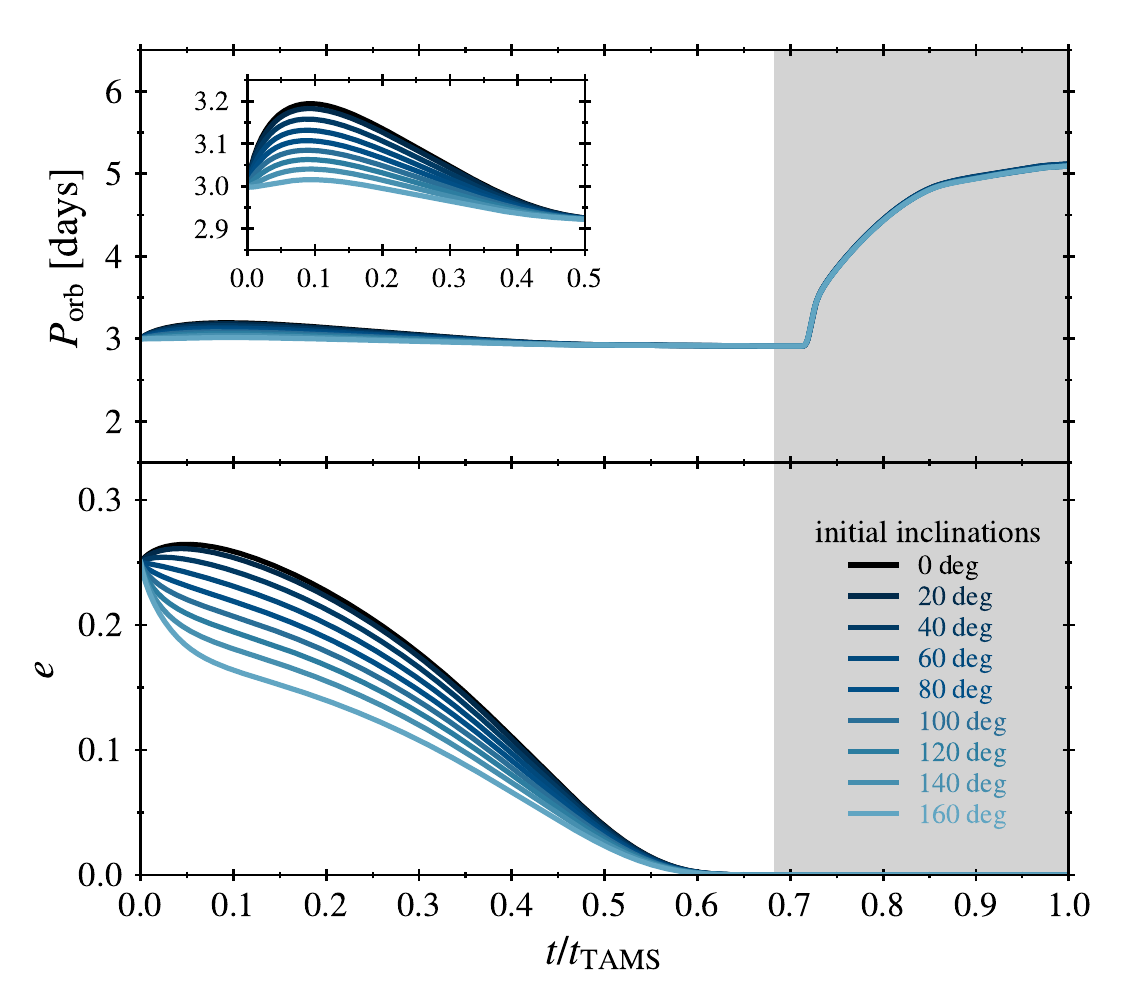}
   \caption{Main sequence evolution of $P_{\rm orb}$ (top panel) and $e$ (bottom panel). Each shade of blue  represents a detailed
      evolution assuming an initial inclination (see  legend, bottom panel). The coplanar case ($i = 0$) is shown in black.
      The zoomed-in region in the top panel presents a clear picture of the different  orbital periods around the first half of the MS
      evolution of the binaries. The grey region indicates the MT phase.}
   \label{fig:orbital_params_evolution_caseA}
\end{figure}

The impact of different initial inclinations on $P_{\rm orb}$ and $e$ is shown in Figure~\ref{fig:orbital_params_evolution_caseA}. It is
clear  that there is  little change in $P_{\rm orb}$ due to the inclination evolution. The most important changes occur at the
beginning of the MS evolution of the star, where $P_{\rm orb}$ increases faster for the smaller values of $i$. However, after half of the
MS evolutionary timescale, the differences in $P_{\rm orb}$ disappear. On the other hand, when studying the change in $e$, we see a clear
difference before reaching a MT phase. In this case, inclinations higher than 90 degrees are able to change the slope of the eccentricity
derivative, leading to the different shapes shown in the bottom panel of Figure~\ref{fig:orbital_params_evolution_caseA}. However, in all
cases the binary tends to reach the equilibrium state characterized by a circular orbit in less than $70\%$ of the MS evolutionary
timescale.

\begin{figure}
   \centering
   \includegraphics[width=\hsize]{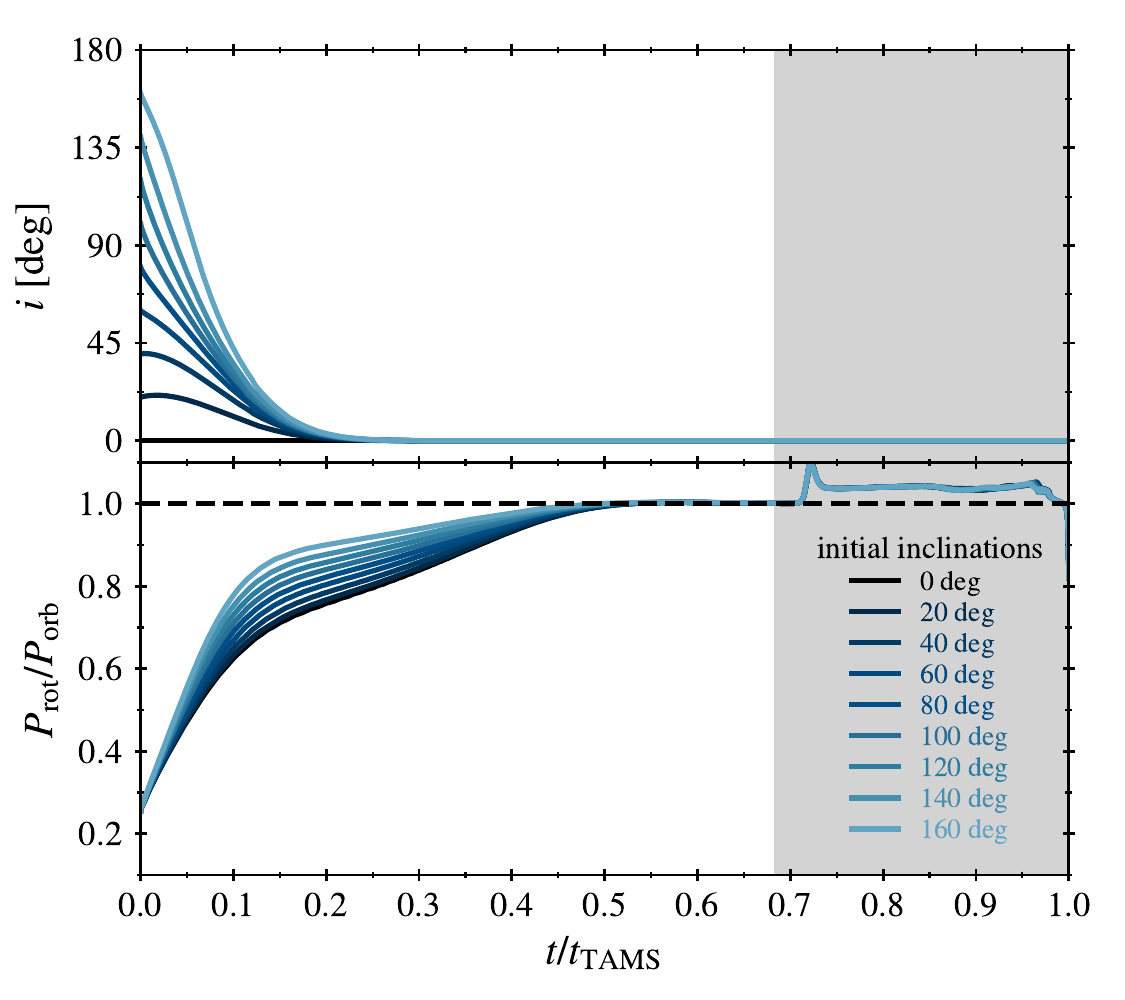}
   \caption{Evolution during main sequence of $i$ (top panel) and $P_{\rm rot}/P_{\rm orb}$ (bottom panel). Each shade of blue  
      represents a detailed evolution assuming an initial inclination (see legend,  bottom panel). The coplanar case
      ($i = 0$) is shown in black. The case  $P_{\rm rot}/P_{\rm orb} = 1$, which represents a synchronous orbit, is shown as a black
      dashed line. The grey region indicates the MT phase.}
   \label{fig:orbital_params2_evolution_caseA}
\end{figure}

We show the evolution of the inclination between the spin of the star and the orbital AM in
Figure~\ref{fig:orbital_params2_evolution_caseA}, which reflects that the slope of $i$ in function of time ($t$) varies according to its
initial value. For all cases, we find that in less than $20\%$ of the MS lifetime of the star the binary reaches coplanarity ($i=0$), after
which the equations derived by \citet{hut1981} are recovered. In addition, in Figure~\ref{fig:orbital_params2_evolution_caseA} we present
the evolution towards a synchronous orbit, by showing the ratio  of the rotational period of the star, $P_{\rm rot}$, to
$P_{\rm orb}$. Before reaching the equilibrium state, the rotational rate differs for different values of $i$, acquiring faster rotations
for the smallest inclinations. Here it is clearly seen how this equilibrium condition is achieved at around half of the MS lifetime, thus
safely allowing the use of the Roche formalism for the MT stage, which  happens at around $70\%$ of the MS lifetime.

From Figs.~\ref{fig:orbital_params_evolution_caseA}~and~\ref{fig:orbital_params2_evolution_caseA} we conclude that the non-coplanar
tides acting on a tight binary in which its components interact during core-H burning push the binary to reach the equilibrium state before
the start of the MT phase that characterizes an accreting HMXB.

\subsection{Mass transfer after core hydrogen depletion}

In order to study the impact of tides over a broad range of conditions, we also explore cases with a wider binary configuration. Here we
present the results for the evolution of a binary with the same component masses, eccentricities, and rotational rates as in
Section~\ref{subsection:results-caseA}, but with a higher $P_{\rm orb}$ of $50$~days. For these cases the beginning of the MT phase occurs
after the end of MS evolution, when the star   crosses the Hertzsprung gap (HG), such that the MT is commonly referred to as case B of MT
\citep{kippenhahn1967}.

\begin{figure}
   \centering
   \includegraphics[width=\hsize]{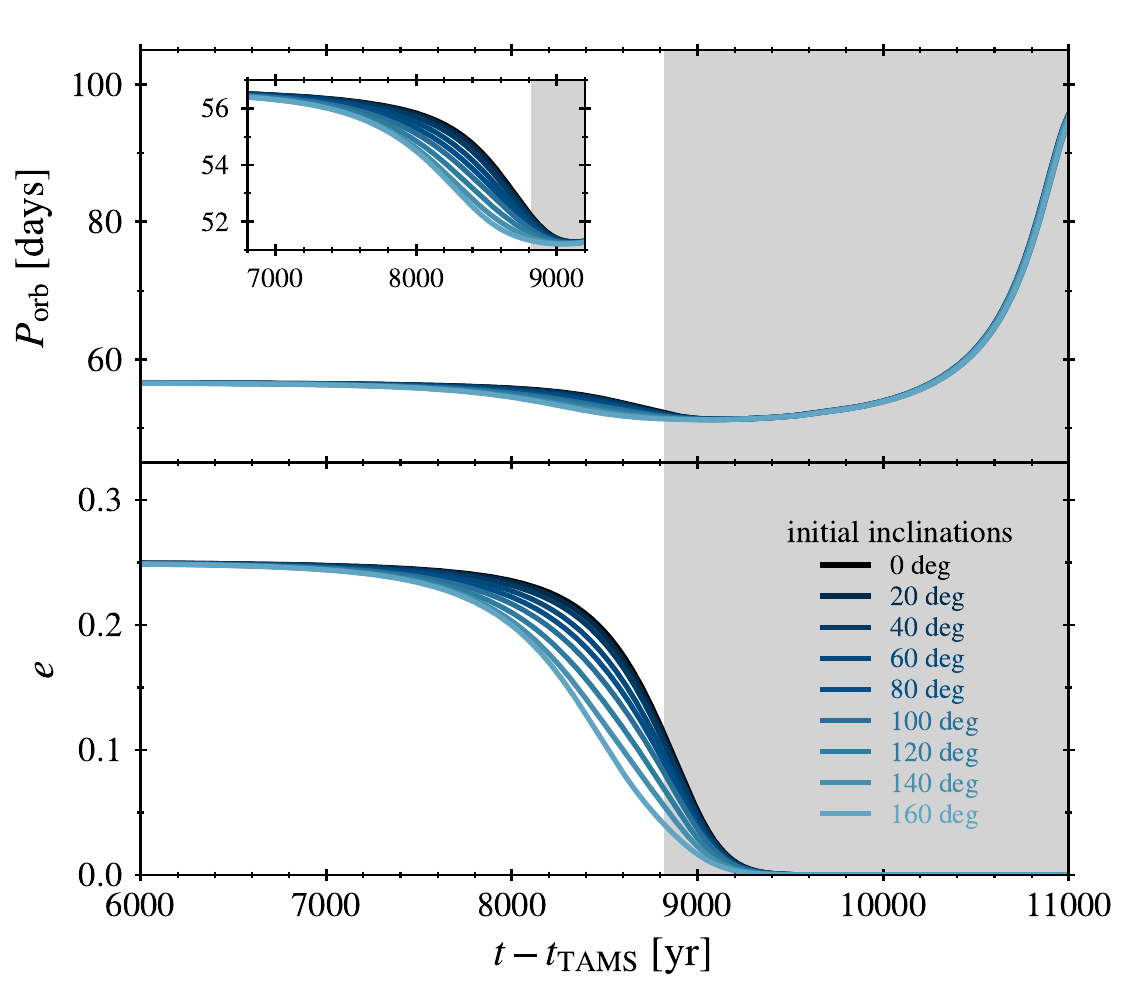}
   \caption{Same parameter evolutions as in Figure~\ref{fig:orbital_params_evolution_caseA}, but   the time axis starts just
      after the end of the MS. The evolution of these parameters before $\sim$6\,000~yr after reaching the end of the MS is the same in all
      the different  inclination cases. In the zoomed-in region  the small differences present in $P_{\rm orb}$ just before the
      beginning of the MT phase are shown. The grey region indicates the MT phase.}
   \label{fig:orbital_params_evolution_caseB}
\end{figure}

The evolution of $P_{\rm orb}$ and $e$ for different values of $i$ are shown in Figure~\ref{fig:orbital_params_evolution_caseB}. Similar to
 case A of MT, $P_{\rm orb}$ behaves in an analogous manner for the different $i$ values. There is, however, a small difference just
before the start of the MT phase, as highlighted in the inset of the figure. For the evolution of $e$, shown in the bottom panel of
Figure~\ref{fig:orbital_params_evolution_caseB}, we see that its progress over time depends on the inclination, with smaller values of $i$
favouring smaller changes towards a circular orbit. Hence, for this case of MT a circular orbit is reached a few hundred years after
the start of the MT phase, around~9\,500~yr after the end of the MS evolution.

\begin{figure}
   \centering
   \includegraphics[width=\hsize]{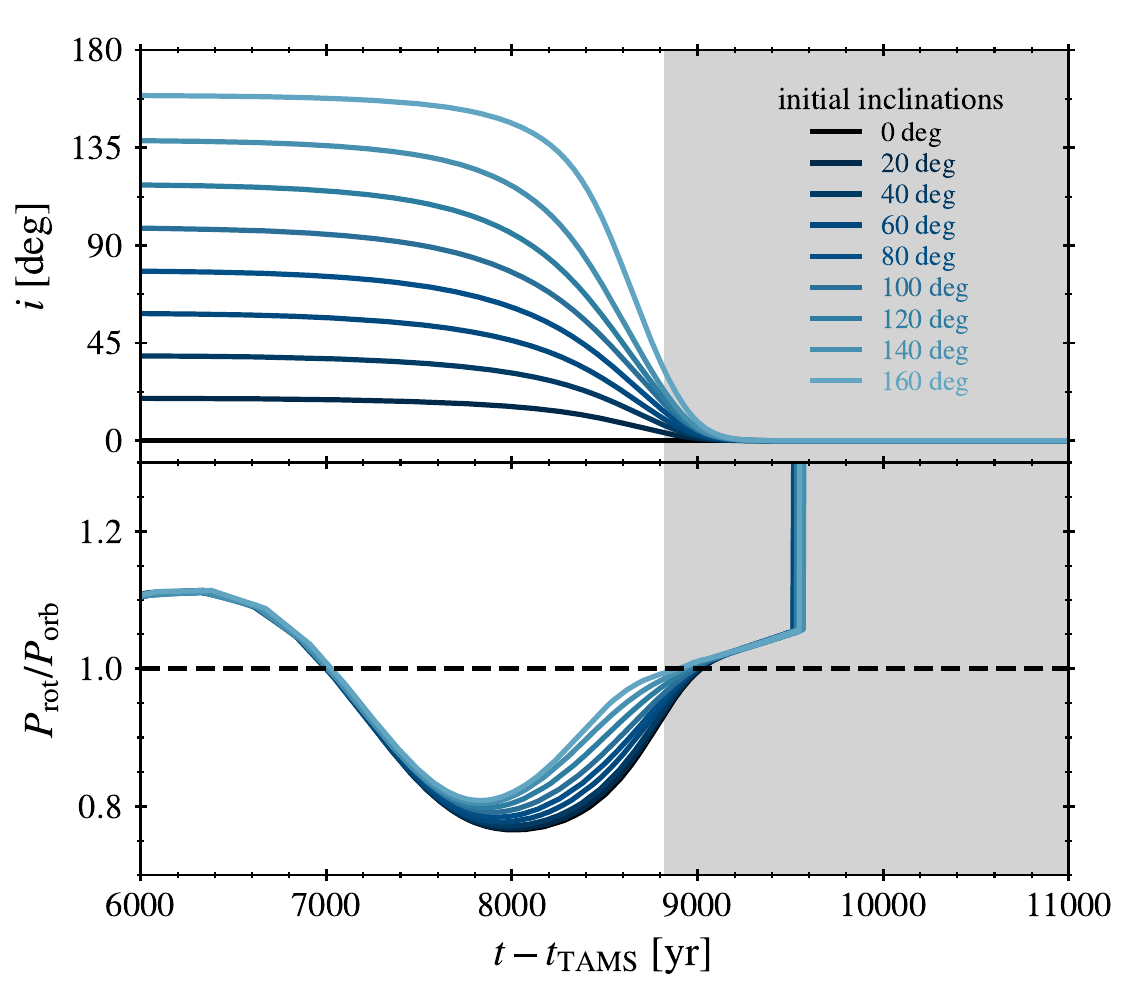}
   \caption{Same parameter evolutions as in Figure~\ref{fig:orbital_params2_evolution_caseA}, but  with a time axis as described in
      Figure~\ref{fig:orbital_params_evolution_caseB}. The black dashed line represents the synchronous orbit.}
   \label{fig:orbital_params2_evolution_caseB}
\end{figure}

Regarding the inclination evolution shown in Figure~\ref{fig:orbital_params2_evolution_caseB}, we see that it starts to change some
thousands of years before the beginning of the MT phase, with the more abrupt changes related to the higher initial $i$ values. This
behaviour is a consequence of the increase in the radius of the star such that the ratio $R_1 / a$ becomes more important in the equations
presented in Section~\ref{subsection:tidal-model}. All these cases tend to a coplanar state that is reached just after the start of the MT
phase.

An important result found for this MT case is connected to the evolution of $P_{\rm rot}/P_{\rm orb}$ (see bottom panel of
Figure~\ref{fig:orbital_params2_evolution_caseB}). On the one hand, we find that $\sim$7\,000~yr after the end of the MS the star begins to
rotate at a faster rate than $P_{\rm orb}$, but this tendency is reverted during the last thousand years before the start of the MT
phase. On the other hand, during the first hundred years of the MT phase, the star does not have a totally synchronous orbit. This equilibrium state is briefly achieved until the clear slow down of the spin of the star occurring around $\sim$9\,500~yr after the MS.

Thus, we can conclude that the different initial values of $i$  reach equilibrium conditions on a similar timescale, but the evolution
towards those conditions can be very different, as in the cases of $e$ and $P_{\rm rot}/P_{\rm orb}$.

\subsection{Different physical assumptions}

\begin{figure}
   \centering
   \includegraphics[width=\hsize]{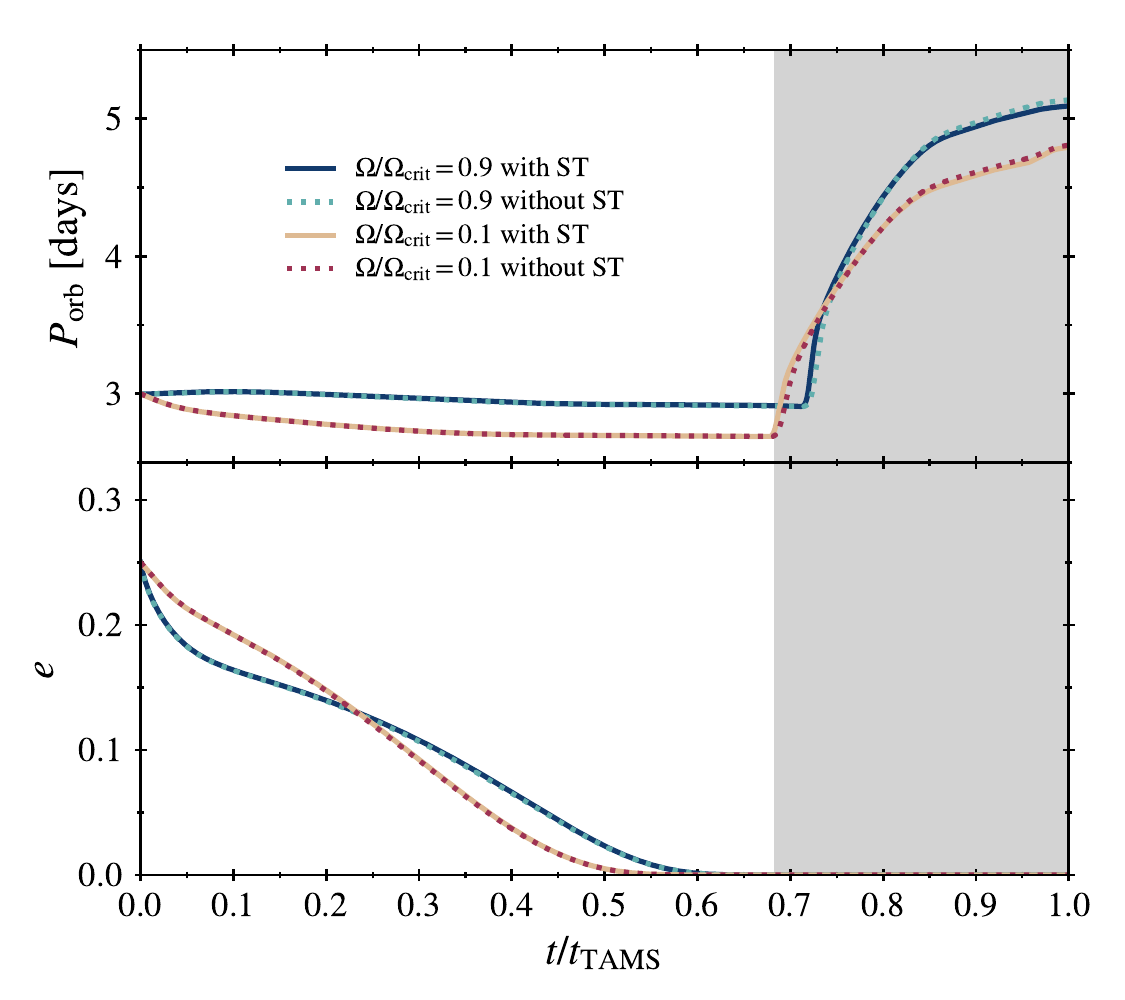}
   \caption{Same evolution as in~Figure~\ref{fig:orbital_params_evolution_caseA}, but for different conditions in the spin of the star and
      the AM transport (see legend, top panel).}
   \label{fig:orbital_params_evolution_caseA_appendix}   
\end{figure}

To explore a wider range of conditions that could modify the evolution of the binaries, we computed a set of models in which the initial
binary conditions are fixed, with masses $M_{\rm BH} = 15$~M$_\odot$ and $M_1 = 20$~M$_\odot$, in an eccentric orbit of $e = 0.25$ and
$P_{\rm orb} = 3$~days and an inclination of $i = 80$~deg. We explore two different initial rotation velocities
$\Omega_1 / \Omega_{\rm crit} = 0.1, 0.9$, and for each of these cases we turn on and off the contribution of the ST dynamo to the AM
transport (respectively labelled in Figure~\ref{fig:orbital_params_evolution_caseA_appendix} as `with ST' and `without ST').

The different results for the evolution of $P_{\rm orb}$ and $e$ are presented in Figure~\ref{fig:orbital_params_evolution_caseA_appendix}.
We see that the main difference arises when the initial spins are changed. This behaviour is maintained even after the MT starts as
$P_{\rm orb}$ is shorter for the initially less rapidly rotating star. The impact of using the ST dynamo seems to be negligible for these
two orbital parameters. In the bottom panel of Figure~\ref{fig:orbital_params_evolution_caseA_appendix} we show that the condition of
circularity is reached first for the slowly rotating case.

\begin{figure}
   \centering
   \includegraphics[width=\hsize]{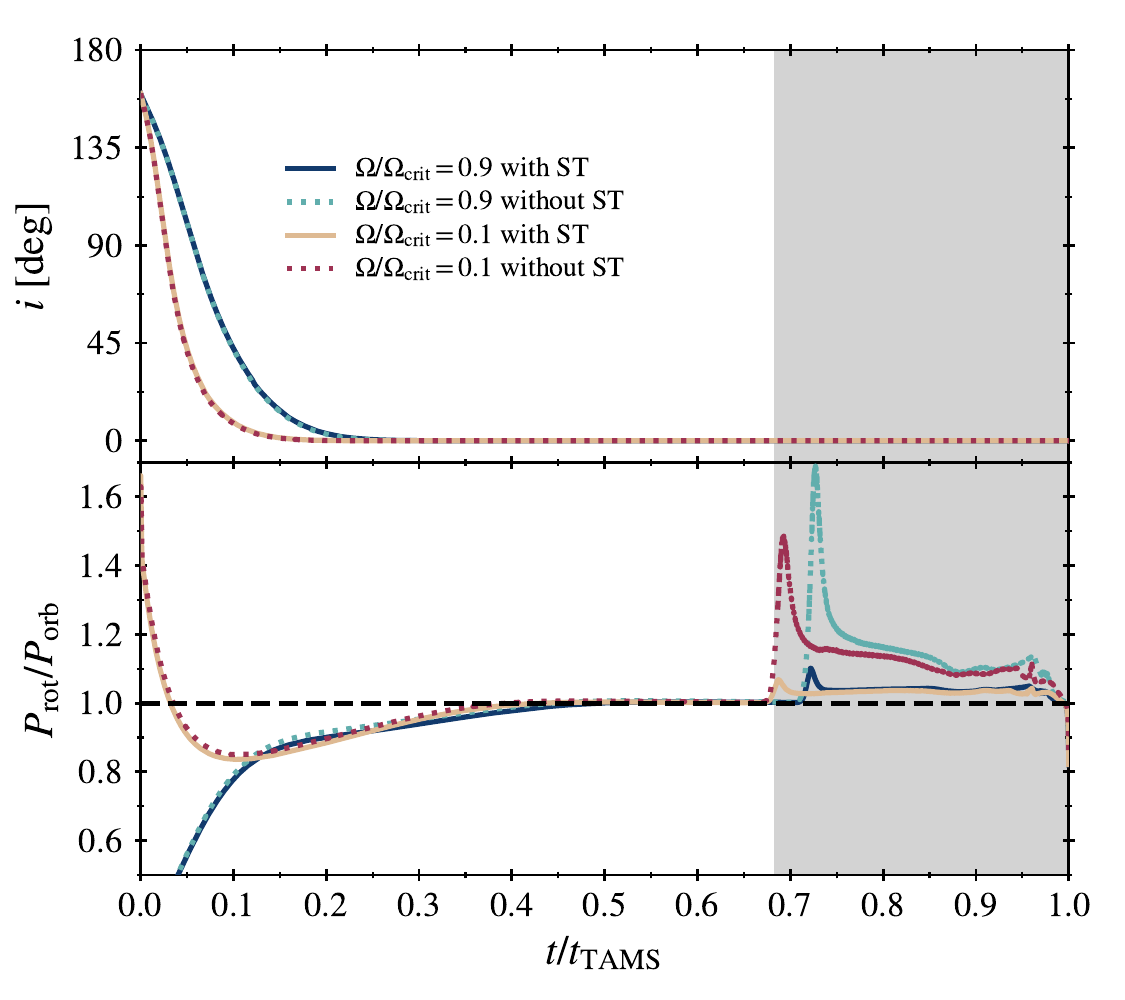}
   \caption{Same evolution as in~Figure~\ref{fig:orbital_params2_evolution_caseA}, but for different conditions in the spin of the star and
      the AM transport (see legend, top panel). The black dashed line represents a synchronous orbit.}
   \label{fig:orbital_params2_evolution_caseA_appendix}
\end{figure}

However, whether the ST dynamo is considered or not can be seen when studying the evolution of the spin of the star. In
Figure~\ref{fig:orbital_params2_evolution_caseA_appendix} we can see this difference in the AM transport becoming important during the MT
phase, as the ratio $P_{\rm rot}/P_{\rm orb}$ shows a clear change in its value when this contribution is turned on and off.

Overall, the impact of the spin of the star is much more important during the evolution prior to the start of the MT phase, as  is
expected, due to the explicit dependence in the tidal equations~(\ref{subsection:tidal-model}). Instead,  the AM transport becomes relevant once
the MT has started, and the binary has reached the equilibrium state.

%--------------------------------------------------------------------
\section{Summary and conclusions}
\label{section:conclusions}

At the end of their evolution, most massive stars  go through a core collapse to form a compact object, which could be a NS or a BH.
This formation is often associated with very powerful astrophysical transients such as supernovae and gamma-ray bursts. Many physical
processes operating during this short-lived stage still remain unknown, one important process being whether this core collapse is symmetric or not.
This dichotomy can be seen in the momentum kick imparted to the newly born object, although the presence of this kick may depend on the
nature of the compact object, with  NSs having well-constrained kick magnitudes \citep{hobbs2005}, while kicks in BHs are more uncertain
\citep{brandt1995,willems2005,wong2012,wong2014,vanbeveren2020,callister2021,stevenson2022}. In addition, if these stars are members of
binary systems, the momentum kick imparted to the remnant will, through linear momentum conservation, modify the orbital parameters of the
binary: $P_{\rm orb}$, $e$, and $i$. After that, the  orbital and internal AM evolution will be dominated by the effect of tides operating
on the binary system.

% \ASB{This dichotomy can be seen in the momentum kick imparted to the newly born object, although the presence of this kick may depend on
% the nature of the compact object, with NSs having well constrained kick magnitudes \citet{hobbs2005} while kicks in BHs being more
% uncertain. \citet{callister2021} argue that large kicks must be imparted to BHs in order to explain the spins measurements of the LIGO
% and Virgo collaboration \citep[although see][for a dicussion on this]{stevenson2022}. Another argument in favour of large kicks to BHs at
% birth is the absence of wind-fed BH HMXBs in the solar neighborhood which might be associated to the disruption of binaries consistent of
% a Wolf-Rayet (WR) and a O-type star when the WR collapses into a BH \citep{vanbeveren2020}. Studies of individual binaries show some
% cases in which large kicks are needed \citep{brandt1995}, while others favouring smaller kicks \citep{willems2005,wong2012,wong2014}.}

In this work we address the topic of how these tides operate on the progenitors of HMXBs in which, after such an asymmetric kick is
imparted, the spin of the star is not aligned with the orbital AM. For this purpose, we include the formulation of \citet{repetto2014} for
arbitrary inclinations in the stellar evolution code {\tt MESA}. With this tool, we study the evolution of the binary parameters
considering different initial values of the inclination between the AM of the star and the orbital plane $i$ for cases A and B of MT
\citep{kippenhahn1967}.

For the evolution through both cases of MT, we note that the most noticeable differences arise in the evolution of $e$, $i$, and
$P_{\rm rot}$. The changes in these parameters are driven by tidal coupling, which forces the system to evolve towards an equilibrium state
characterized by circularization, coplanarity, and synchronous orbit. This state is achieved faster for the more compact binary (i.e. the
conditions of  case A of MT) as a consequence of the strong dependence of tides with the separation between the binary members (see
Section~\ref{subsection:tidal-model}).

Furthermore,  in this work we explored the role of the initial spin on the subsequent binary evolution. We find the spin to be an important
variable that regulates the timescale for each of the different conditions of the equilibrium state. We also study an important yet
debated mechanism of AM transport: the ST dynamo. As already shown by other authors \citep{qin2019} its absence could help to explain the
high spin measured in the BHs of HMXBs, although this might not be consistent with the population of BHs detected by the LIGO and Virgo
collaboration \citep{abbott2021}. In this work we find that the role of this mechanism is secondary when analysing the change on the
binary parameters due to tidal evolution, although it becomes important once a binary reaches the onset of a MT phase. However, the ST dynamo
might play an important role in scenarios where a rapidly rotating core of a star is needed at the end of the evolution to reproduce
energetic transient events \citep{fuller2022}.

Many physical uncertainties could affect our results across this modelling. Understanding the impact of such limitations is a key factor.
On the one hand, the set of non-linear differential equations that describe the evolution of the semi-major axis, eccentricity, rotation
rate, and inclination of the rotation axis with respect to the orbital plane, as presented by \citet{repetto2014}, is considered until the
onset of a MT phase. Once a star is near to filling its Roche lobe, the quadrupolar terms in the perturbation theory of tides of
\citet{hut1981} might stop being a good approximation, with tides becoming stronger than predicted. Moreover, the Roche formalism requires
synchronous orbit between the components of the binary and $P_{\rm orb}$; in our case B of MT, we find small deviations of the synchronous
orbit state at the beginning of the MT phase which is rapidly achieved during the first years of MT. Lastly, the transport of AM is also
uncertain and will certainly have an impact on the results during the MT phase. There is observational evidence that this transport is
highly efficient such that most stars end their evolution with a smaller content of AM than predicted
\citep{cantiello2014,fuller2014,spada2016,aerts2019,fuller2019,eggenberger2019,denhartogh2019}. Although this is particularly true for
measurements of BHs with low spin values \citep{abbott2021}, there seem to be HMXBs containing BHs with high spin values
\citep{miller2015}. This issue is currently under debate \citep{qin2019,bavera2020,fishbach2021,belczynski2021}.

The results of this study show that prior to the X-ray binary phase, the evolution of the orbital parameters   differ according to the
value of the inclination of the rotational axis with respect to the orbital plane. Thus, for some binaries, matching observed quantities
with those coming from stellar evolution simulations might help constrain the conditions at core collapse, as well as the momentum kick
imparted to the newly born compact object.

{\em Software:} {\tt MESA}\footnote{\url{http://mesa.sourceforge.net/}} \citep{paxton2011,paxton2013,paxton2015,paxton2018,paxton2019},
{\tt ipython/jupyter} \citep{jupyter}, {\tt matplotlib} \citep{hunter2007}, {\tt numpy} \citep{harris2020}.

%--------------------------------------------------------------------
\begin{acknowledgements}

ASB is a fellow of CONICET. FG and JAC are CONICET researchers. FG and JAC acknowledge support by PIP 0113 (CONICET). JAC is a Mar\'ia
Zambrano researcher fellow funded by the European Union -NextGenerationEU- (UJAR02MZ). This work received financial support from
PICT-2017-2865 (ANPCyT). JAC was also supported by grant PID2019-105510GB-C32/AEI/10.13039/501100011033 from the Agencia Estatal de
Investigaci\'on of the Spanish Ministerio de Ciencia, Innovaci\'on y Universidades, and by Consejer\'ia de Econom\'ia, Innovaci\'on,
Ciencia y Empleo of Junta de Andaluc\'ia as research group FQM- 322, as well as FEDER funds. SC acknowledges the CNES (Centre National
d’Etudes Spatiales) for the funding of MINE (Multi-wavelength INTEGRAL Network). ASB, FG, FF and SC are grateful to the LabEx UnivEarthS
for the funding of Interface project I10 « From binary evolution towards merging of compact objects ».

\end{acknowledgements}

\bibliographystyle{aa}
\bibliography{aanda}

\end{document}